\documentclass{ws-mpla}
\usepackage[super]{cite}
\usepackage{graphicx}
\usepackage{url}
\newcommand*{\pT}{\ensuremath{p_\text{T}}}

\begin{document}

\markboth{Y.C. Yap}
{Recent observation and measurements of diboson processes from the ATLAS experiment}

\catchline{}{}{}{}{}

\title{Recent observation and measurements of diboson processes from the ATLAS experiment}

\author{\footnotesize Yee Chinn Yap}

\address{DESY, Notkestr. 85, 22607 Hamburg, Germany\\
yee.chinn.yap@cern.ch}

\maketitle

\pub{Received (Day Month Year)}{Revised (Day Month Year)}

\begin{abstract}

This review covers results at a centre-of-mass energy of $\sqrt{s}=13$~TeV from the ATLAS experiment that have been published, or submitted for publication, up to April 2020. It summarizes results on the inclusive production cross-section measurements of boson pairs and of the electroweak production of diboson in association with two jets. The measurements either use the full integrated luminosity of 139~$\text{fb}^\text{-1}$ collected by the ATLAS detector at the LHC from 2015 to 2018, or a partial dataset of 36 ~$\text{fb}^\text{-1}$. The inclusive production rates of diboson are studied to high precision. These measurements provide stringent tests of the electroweak sector of the Standard Model and allow search for new physics via anomalous triple and quartic gauge boson couplings.

\keywords{Electroweak; LHC; Diboson; ATLAS.}
\end{abstract}

\ccode{PACS Nos.: include PACS Nos.}

\section{Introduction}
The study of diboson production at the LHC plays an important role in tests of the electroweak (EW) sector of the Standard Model (SM) and searches for new physics at the TeV scale. Diboson production is also connected to the spontaneous breaking of the EW gauge symmetry. Not only that the Higgs boson can decay into gauge boson pairs, but its presence is also necessary to restore the unitarity in the amplitude of the longitudinal gauge boson scattering process \cite{VBSunitarity1, VBSunitarity2}. 

Diboson processes and the scattering of two vector bosons are sensitive to triple and quartic gauge boson couplings. A broad range of beyond SM phenomena result in anomalous gauge boson couplings and can be probed in diboson processes. Diboson or diboson in association of two jets are signatures of many new physics models thus precise measurements are important to constrain their contributions as background in searches or in the study of the Higgs boson.

Furthermore, perturbative QCD (pQCD) and next-to-leading-order (NLO) EW corrections can be tested by studying diboson production which is sensitive to such higher-order corrections.
 
From 2015 to 2018, a period known as Run 2, the LHC operated at a centre-of-mass energy of $\sqrt{s}=13$~TeV. This article presents several results obtained using data collected by the ATLAS experiment during this period.

\section{Dibosons in the Standard Model}

\begin{figure}[h]
\centerline{\includegraphics[width=4.0in]{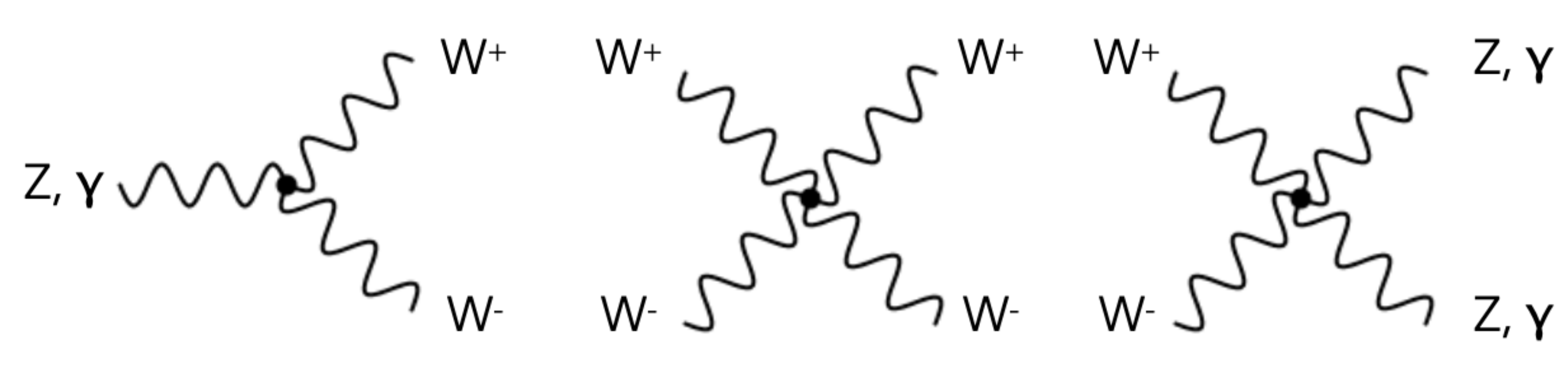}}
\vspace*{8pt}
\caption{Allowed gauge boson self-interactions in the SM.}
\protect\label{fig:vertices}
\end{figure}

In the SM, certain triple and quartic gauge boson self-interactions are allowed as a consequence of the  non-Abelian $SU(2)_L \otimes U(1)_Y$ gauge symmetry of the electroweak sector \cite{APich}. The allowed gauge boson self-interactions are shown in Figure \ref{fig:vertices}.

At leading-order (LO), the production of $VV$ with $V=W/Z/\gamma$ proceeds through a $t$-channel process with a quark and an anti-quark initial state. The $s$-channel production is forbidden at the lowest order for $ZZ$, $Z\gamma$ and $\gamma\gamma$ but is allowed for $WW$, $W\gamma$ and $WZ$ via the $WWZ/WW\gamma^*$ vertex. 

The earlier $W^\pm W^\mp$ cross-section measurements at $\sqrt{s} = 7$ and $\sqrt{s} = 8$~TeV by ATLAS \cite{WW7TeV, prelimWW8TeV} and CMS \cite{WW7TeVCMS, prelimWW8TeVCMS} were 15-20\% higher than the theory prediction at the time which was only at NLO in pQCD. The inclusion of higher-order corrections \cite{NNLOWW} has since resolved this discrepancy in the integrated cross-section. Subsequently, the incompatibility with NLO prediction has also been observed in e.g. $WZ$ \cite{WZ8TeV, WZ}. 

The diboson production via gluon fusion arises at NNLO and includes two sub-processes, one involving a fermion loop and the other a virtual Higgs boson. Although the gluon-initiated process only appears at $O(\alpha_\text{S}^2)$, its contribution can be sizeable due to the large gluon flux at the LHC. The two sub-processes $gg \rightarrow VV$ and $gg \rightarrow H \rightarrow VV$ interfere with one another.  

The diboson production cross-sections are known to NNLO in pQCD and NLO in EW. Gluon-initiated production cross-sections are known to $O(\alpha_\text{S}^3)$, or NLO for $WW$ \cite{WWggNLO}, $\gamma\gamma$ \cite{yyggNLO} and $ZZ$ \cite{ZZggNLO}.

The scattering of two vector bosons, $VV\rightarrow VV$, is known as vector boson scattering (VBS). At the LHC, VBS occurs when two vector bosons are radiated from the initial-state quarks in the colliding protons, and then scatter into another pair of vector bosons in the final state. VBS processes involve quartic gauge-boson self-interactions, and the $s$- and $t$-channel exchanges of a gauge or Higgs boson. 

It is an important process to study the mechanism of EW symmetry breaking (EWSB). The amplitude of longitudinal gauge boson scattering diverges at high energies and the presence of the Higgs boson precisely cancels the divergence and restores unitarity at the TeV scale. This cancellation is delicate and any deviation in the SM coupling of the Higgs boson to the gauge boson could break it. Thus the study of VBS is complementary to the direct measurements of the Higgs boson properties to probe the exact nature of EWSB. The study of VBS also allows tests for the presence of anomalous quartic gauge couplings \cite{VBSQGC}.

$VVjj$ final state can be produced not only in VBS processes. The inclusive $VVjj$ production can be split into two categories: \textit{QCD} and \textit{EW}. QCD production of $VVjj$ involves both strong and electroweak interactions, and typically has a much higher cross-section. The purely electroweak production of $VVjj$ consists of a range of processes that include VBS. One cannot study VBS diagrams independently from the other EW processes \cite{VBSensemble}. The contributions from the non-VBS processes could, however, be suppressed with certain kinematic selections.

\section{Experimental aspects} 
\subsection{The ATLAS detector}

The ATLAS detector \cite{PERF-2007-01} is one of the multi-purpose particle detectors at the LHC. It has a forward-backward symmetric cylindrical geometry with nearly 4$\pi$ solid angle coverage. The ATLAS detector is composed of an inner tracking detector (ID) surrounded by a superconducting solenoid providing a 2 T axial magnetic field, electromagnetic and hadronic calorimeters, and a muon spectrometer (MS).

\subsection{Data and simulation}

The analyses that are discussed here include at least one $W$ or $Z$ boson which decays leptonically to electrons or muons. The decay products are used to reconstruct the $W$ and $Z$ bosons.

Events used in the analyses are selected via a trigger system. Single-electron and single-muon triggers are used in all the analyses presented here, with the exception of EW $Z\gamma jj$ and $ZZjj$ analyses which make use of both single- and multi-lepton triggers.
 
The measurements either use the full integrated luminosity of 139~$\text{fb}^\text{-1}$ collected by the ATLAS detector at the LHC from 2015 to 2018, or a partial dataset of 36 ~$\text{fb}^\text{-1}$.
The average number of inelastic $pp$ interactions produced per bunch crossing for the full run 2 dataset is $<\mu>=33.7$, while it is 23.7 for the partial 2015-2016 dataset.

Monte Carlo (MC) simulation is used to model the signal and various background processes. The simulated samples were produced with various MC event generators and processed through a full ATLAS detector simulation \cite{ATLASsim} based on \textsc{Geant4} \cite{GEANT4} and are reconstructed using the same algorithms as used for data. Additional $pp$ interactions (pile-up) were modelled by overlaying each MC event with minimum-bias events. Simulated events were then reweighted to match the distribution of the average number of interactions per bunch crossing observed in data.

\subsection{Physics objects}

Electron candidates are reconstructed from clusters of energy deposits in the electromagnetic calorimeter with information about charged tracks reconstructed in the ID. They are required to pass certain transverse momentum \pT~threshold and to be located within the pseudorapidity range $|\eta|<2.47$ (excluding $1.37<|\eta|<1.52$). 
 
The reconstruction of photon candidates is similar to the electron. A fraction of photons converts into electron-positron pairs within the ID. They are classified as converted if the photon cluster is matched to conversion track(s), and otherwise as unconverted. Photon clusters are required to have a pseudorapidity in the range $|\eta|<2.37$, excluding $1.37<|\eta|<1.52$.

Muon candidates are reconstructed by matching tracks from the MS to a corresponding track in the ID. The muon momentum  is calculated by combining the MS measurement, corrected for energy loss measured by the calorimeter, and the ID measurement. Muon candidates are reconstructed within $|\eta|<2.7$, while the ID only covers $|\eta|<2.5$. Some analyses limit the muon pseudorapidity range to $|\eta|<2.5$ to use only combined muon candidates, where tracks are reconstructed in both the ID and the MS.

To ensure that candidate electrons and muons originate from the primary interaction vertex, they are required to have small longitudinal and transverse impact parameters. Each primary vertex candidate is reconstructed from at least two associated tracks with $\pT>0.4$~GeV, and the one with the highest sum of the squared transverse momenta of its associated tracks is selected.

For photons, electrons and muons, additional identification requirements are imposed. The objects are also usually required to be isolated using tracks and calorimetric information. 

Jets are reconstructed from topological clusters of calorimeter energy deposits using the anti-$k_t$ \cite{antikt, fastjet} algorithm with radius parameter of $R = 0.4$. Pile-up jets in the ID acceptance are suppressed using a multivariate combination of the track and vertex information, also known as a jet-vertex-tagger \cite{jvt}. In addition, jets containing $b$-hadrons ($b$-jets) are identified in the ID volume using a multivariate algorithm \cite{btagging}.

The missing transverse momentum $E_\text{T}^\text{miss}$ is computed as the negative of the vectorial sum of the transverse momenta of all the charged leptons and jets, as well as the tracks originating from the primary vertex but not associated with any of the leptons or jets \cite{met}.

\subsection{Cross section measurements}

The cross-section of a process is often measured in a fiducial phase space region defined by particle-level requirements similar to those at reconstruction level. The fiducial volume at particle level is defined using stable particles (defined as having a mean lifetime $c\tau > 10$~mm). For electrons and muons, QED final-state radiation is partly recovered by adding the four momenta of prompt photons within a cone of size $\Delta R = 0.1$ around the lepton to the lepton four momentum. The corrected leptons are known as dressed leptons. Particle-level jets are built with the anti-$k_t$ algorithm with radius parameter $R = 0.4$, using final-state particles as input.

The integrated cross-section in the fiducial phase-space region is calculated as

\begin{equation}
  \sigma^\text{fid} = \frac{N_{obs} - N_{bkg}}{C \times \mathcal{L}} \,,
\end{equation}

where $N_{obs}$ is the observed number of selected events in the data in the signal region, $N_{bkg}$ is the estimated number of background events, $\mathcal{L}$ is the integrated luminosity of the analysed dataset, and the correction factor $C$ corrects for detection efficiency and acceptance. $C$ is defined as the ratio of the number of selected events in the signal region at reconstruction level to the number of events in the fiducial phase space at particle level.

The total cross-section is obtained as  
\begin{equation}
  \sigma^\text{total} = \sigma^\text{fid} / A, 
\end{equation}

where $A$ is the fiducial acceptance calculated as the ratio of the number of events in the fiducial phase space to the number in the total phase space.

The differential cross-sections are extracted using an unfolding procedure to correct for inefficiencies and resolution effects. An iterative Bayesian unfolding method \cite{iterativeBayesian} is used in $Z\gamma$, $ZZ\rightarrow\ell\ell\nu\nu$ and $WW$ cross-section measurements presented in this article.

\section{Inclusive diboson measurements}

\begin{figure}[h]
\centerline{\includegraphics[width=4.0in]{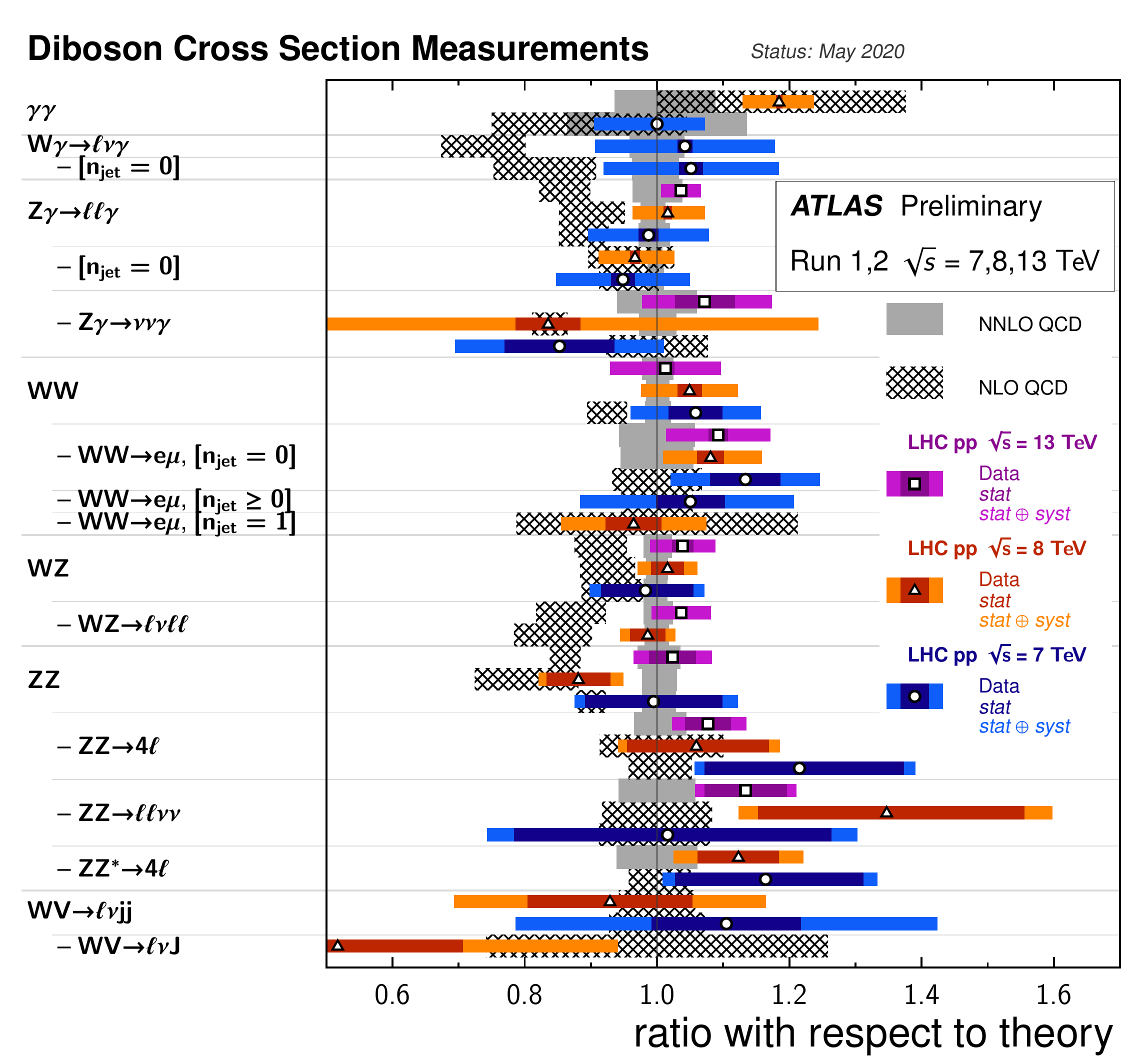}}
\vspace*{8pt}
\caption{The data/theory ratio for several diboson fiducial production cross-section measurements in ATLAS, corrected for branching fractions. Reproduced from Ref.~\protect\refcite{ATL-PHYS-PUB-2020-010}.}
\protect\label{fig:ATLASdibosonsummary}
\end{figure}

Measurements of diboson processes are carried out by ATLAS in all possible bosonic final states, as summarized in Fig.~\ref{fig:ATLASdibosonsummary}. In most cases, the cross-sections are compared to theoretical prediction at NNLO and NLO QCD. The data shows generally good agreement with the NNLO prediction and is above the NLO expectation.  
 
Measurements at $\sqrt{s} = 13$~TeV of all diboson processes have been published apart from $W\gamma$ and $\gamma\gamma$ processes.

\subsection{$Z(\rightarrow\ell^+ \ell^-)\gamma$}

The analysis \cite{inclusiveZy} uses a data sample with an integrated luminosity of 139~$\text{fb}^{-1}$ collected from 2015 to 2018. The production cross-section for the process $pp \rightarrow \ell^+ \ell^- \gamma + X (\ell=e, \mu)$ is measured within a fiducial phase-space region. Candidate $\ell^+ \ell^- \gamma$ events are selected by requiring a photon together with an opposite-charge, same-flavour lepton pair. 

$Z\gamma$ production cross-section is also measured in $\nu\bar{\nu}\gamma$ \cite{nunugamma} and $b\bar{b}\gamma$ \cite{bbgamma} channels using 36~$\text{fb}^{-1}$ data in phase space regions with high photon \pT. The $\nu\bar{\nu}\gamma$ channel has better sensitivity \cite{Zgamma8TeV} on anomalous $ZZ\gamma$ and $ZZ\gamma$ couplings while $\ell^+ \ell^- \gamma$ channel allows cross-section measurements to be made over a wider range of photon transverse energy $E_\text{T}^\gamma$ with lower background and better precision.
 
Table \ref{table:ZgFidReg} shows the definition of the particle-level fiducial phase-space region for this analysis. Photon isolation at particle level is imposed by requiring the scalar sum of the transverse energy of all stable particles (except neutrinos and muons) within a cone of size $\Delta R = 0.2$ around the photon, $E_\text{T}^{cone0.2}$, to be less than 7\% of $E_\text{T}^\gamma$. The sum, $m(\ell\ell) + m(\ell\ell\gamma)$, of the invariant masses of the lepton pair and the $\ell^+ \ell^- \gamma$ system is required to be greater than 182~GeV to ensure that the measurement is dominated by events in which the photon is emitted as initial state radiation rather than from a final state lepton.

\begin{table}[h]
\tbl{Definition of the $\ell^+ \ell^- \gamma$ particle-level fiducial phase-space region.}
{\begin{tabular}{@{}ccc@{}} \toprule
     Photons  & & Electrons/Muons  \\
\colrule
    $E_\text{T}^\gamma > 30$~GeV            & & $\pT^\ell > 30, 25$~GeV  \\
    $|\eta^\gamma| < 2.37$                 & & $|\eta^\ell| < 2.47$  \\
    $E_\text{T}^{cone0.2}>E_\text{T}^\gamma < 0.07$           & & dressed leptons  \\
    $\Delta R(\ell,\gamma) > 0.4$   & &  ~~  \\
\toprule
    \multicolumn{3}{c}{Event selection}  \\
\colrule
    \multicolumn{3}{c}{$m(\ell\ell) > 40$~GeV}  \\
    \multicolumn{3}{c}{$m(\ell\ell) + m(\ell\ell\gamma) > 182$~GeV}  \\
\botrule
\end{tabular}
\label{table:ZgFidReg} }
\end{table}

The dominant background source ($\approx10$\%) originates from $Z+jets$ production in which a jet is misidentified as a photon. Other background contributions arise from top quark ($\approx4$\%) or multiboson production ($\approx~1$\%), and from pile-up background ($\approx 2-3$\%) in which the selected photon and the selected lepton pair originate from different $pp$ interactions within the same LHC bunch crossing. While leptons are required to originate from the primary vertex, no explicit requirement is imposed on the selected photon, hence this background can have a non-negligible contribution since the level of pile-up in this dataset is rather high at $<\mu>=33.7$. The pile-up background is estimated from the data distribution of the longitudinal separation between the reconstructed primary vertex position and the position of the reconstructed photon after extrapolation to the beam-axis.

The differential and integrated cross-section measurements measured in the electron and muon channels are consistent within the uncorrelated uncertainties, and are averaged. Differential cross-sections are measured as functions of $E_\text{T}^\gamma$, $|\eta^\gamma|$, $m(\ell\ell\gamma)$, the angle between the transverse directions of the dilepton system and the photon $\Delta\phi(\ell\ell,\gamma)$, the transverse momentum of the of the $\ell^+ \ell^- \gamma$ system $\pT^{\ell\ell\gamma}$ and the ratio $\pT^{\ell\ell\gamma}/m(\ell\ell\gamma)$.

The measurements are compared with SM predictions obtained from parton-level calculations, corrected to particle level, at NLO and NNLO as well as with predictions from parton shower MC event generators with LO and NLO matrix elements. The effect of NLO EW corrections on the predictions at NNLO is also considered. A small ($\approx1$\%) contribution from the EW production of $Z\gamma jj$ is included in the prediction.

\begin{figure}[h]
\centerline{\includegraphics[width=2.2in]{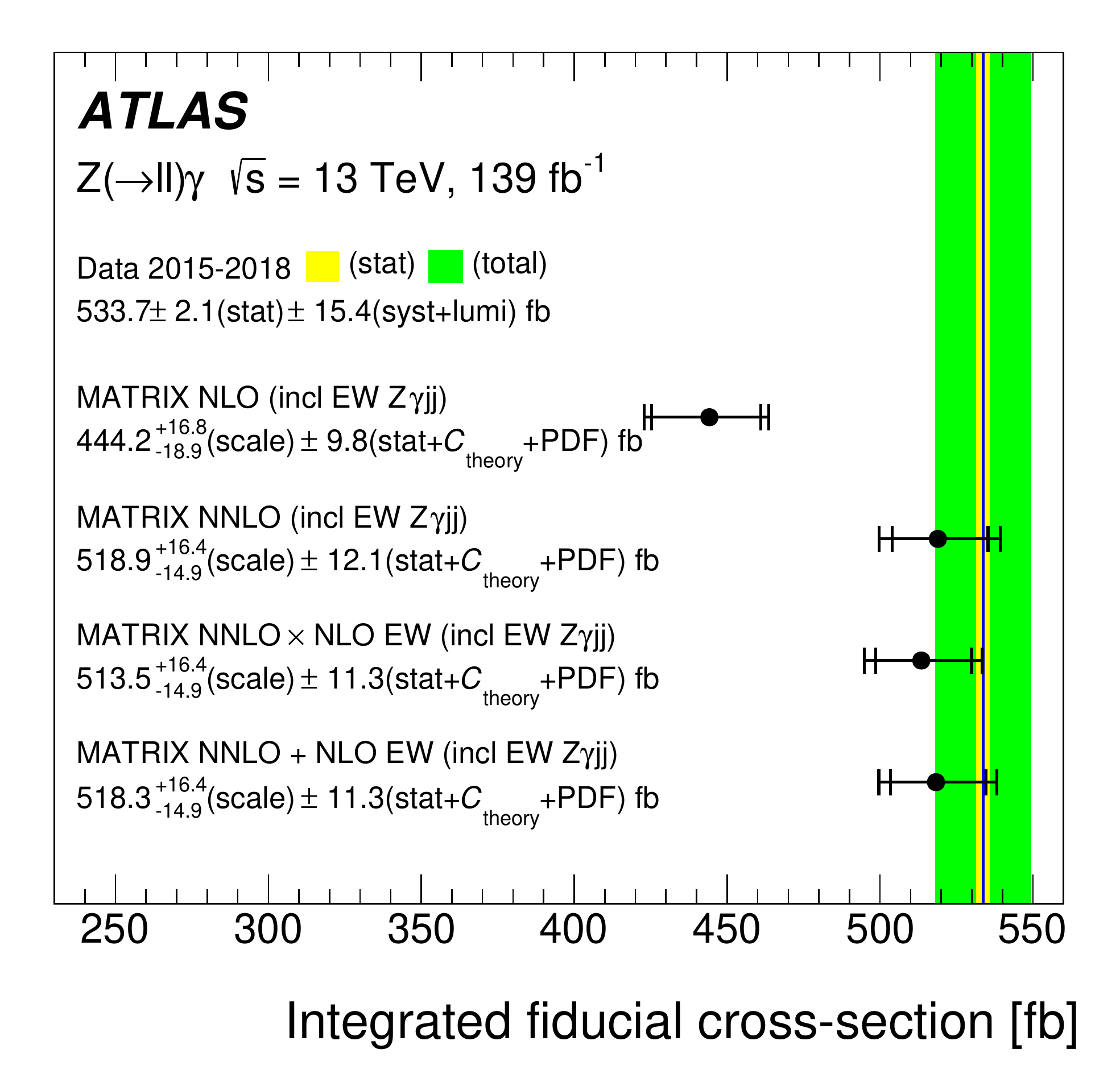} \includegraphics[width=3.1in]{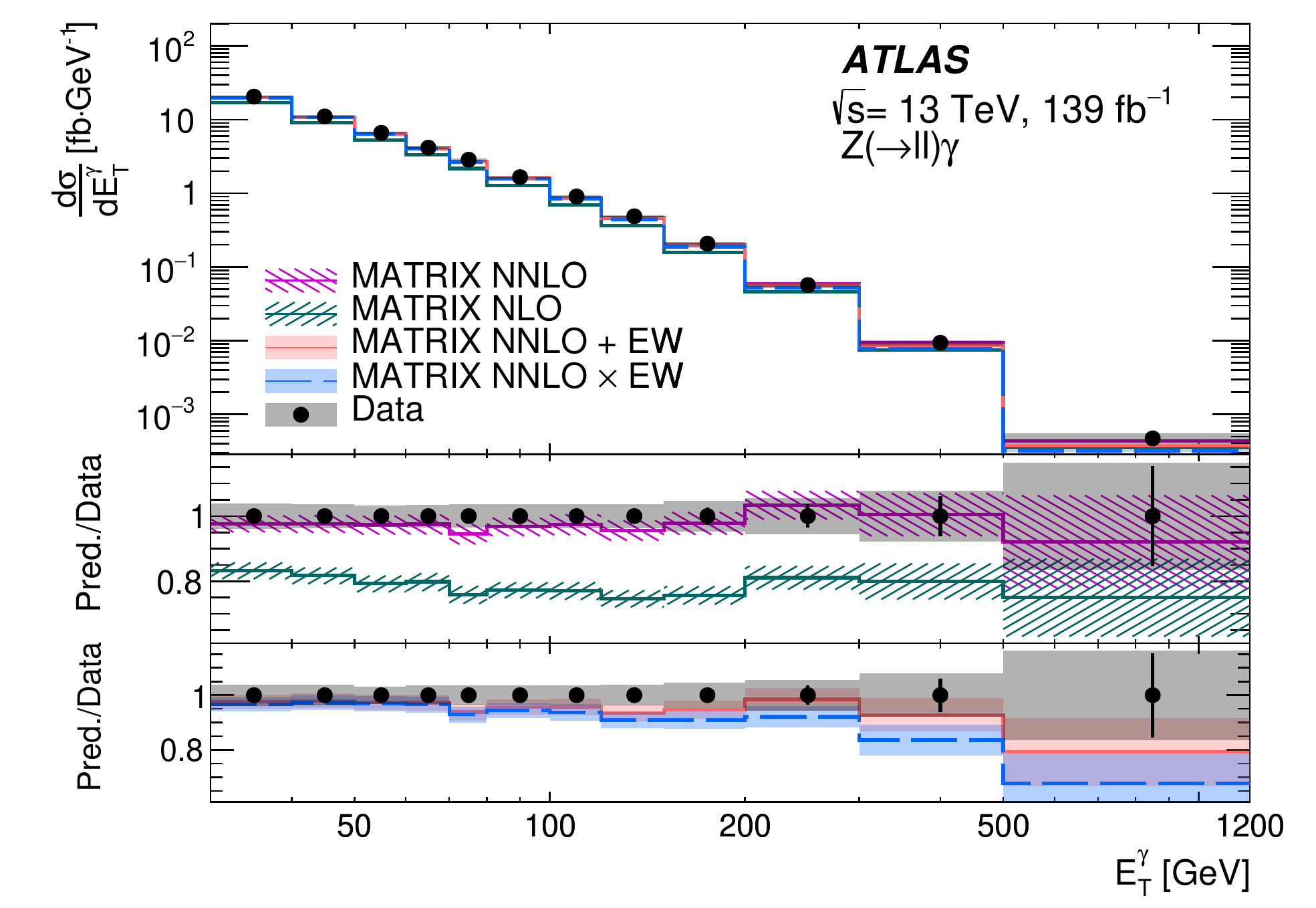}}
\vspace*{8pt}
\caption{Comparison of the measured $Z\gamma$ fiducial cross-section with theoretical predictions from the \textsc{Matrix} generator at NLO and NNLO. The contribution from EW $Z\gamma jj$ production is included in all predictions. The NNLO cross-sections with NLO EW radiative corrections applied multiplicatively and additively are also shown. Reproduced from Ref.~\protect\refcite{inclusiveZy}.}
\protect\label{fig:Zy}
\end{figure}

Fig.~\ref{fig:Zy} shows the comparison of the measured integrated and differential cross-section with theoretical predictions from the \textsc{Matrix} generator  \cite{MATRIX}. The NNLO correction is about +17\% and is significantly larger than the scale uncertainty estimated at NLO. The effect of NLO EW corrections on the predicted differential cross-sections is large and negative, and is different depending on whether the EW corrections are applied multiplicatively or additively. 

\subsection{$ZZ\rightarrow\ell^+ \ell^- \nu \bar{\nu}$}

The measurement of $ZZ$ production in the $\ell\ell\nu\nu$ final state \cite{ZZllvv} is carried out using data collected during 2015 and 2016. Events are selected by requiring a pair of high-$\pT$ isolated opposite sign leptons and significant missing transverse momentum. 

The measurement in the competing $4\ell$ channel has also been performed \cite{ZZ4l}, achieving a precision of 5\%. The $\ell\ell\nu\nu$ final state has higher branching fraction and is more sensitive to anomalous triple gauge couplings (aTGCs). This final state suffers nevertheless from larger background contamination, and stringent experimental selection that requires one $Z$ boson boosted against the other in the transverse plane is needed to keep background at a more manageable level. A signal-to-background ratio of about 1.7 is found after event selection.

The integrated cross-section of $ZZ$ production is measured in a fiducial phase space and then extrapolated to a total phase space. The fiducial phase space is similar to the reconstruction level selection and is defined using dressed leptons. The total phase space is defined using Born level leptons and with the mass requirement $66<m(\ell\ell), m(\nu\nu)<116$~GeV.

The cross-section predictions for the total phase space are corrected for the branching fraction of the $ZZ\rightarrow \ell\ell\nu\nu$ decay. The integrated cross-sections are determined by binned maximum-likelihood fits to the $E_\text{T}^\text{miss}$ distributions. The fiducial cross-section of the combined $ee+\mu\mu$ channels is measured to be $\sigma^\text{fid}_{ZZ\rightarrow\ell\ell\nu\nu} = 25.4 \pm 1.4 (\text{stat.}) \pm 0.9 (\text{syst.}) \pm 0.5 (\text{lumi.})$~fb, in agreement with the SM prediction of $22.4 \pm 1.3$~fb at an accuracy of NNLO in QCD, NLO in EW and NLO QCD for the gluon-gluon initiated production. The total cross-section measured is $17.8\pm1.0(\text{stat.})\pm0.7(\text{syst.})\pm0.4(\text{lumi.})$~fb, compared to $15.7\pm0.7$~fb predicted. The total precision achieved of 7\% is significantly improved over the previous measurement \cite{ZZ8TeV}.

Differential cross-sections are reported in the fiducial region for eight kinematic variables: the transverse momentum of the leading lepton $\pT^{\ell 1}$, the leading jet $\pT^{jet1}$, the dilepton system $\pT^{\ell\ell}$ and the $ZZ$ system $\pT^{ZZ}$, the invariant mass of the $ZZ$ system $m_\text{T}^{ZZ}$, the absolute rapidity of the dilepton system $|\text{y}_{\ell\ell|}$, the azimuthal angle difference between the two leptons $\Delta\phi(\ell_1,\ell_2)$ and the number of jets $N_{jets}$.

The search for aTGCs is carried out using the unfolded $\pT^{\ell\ell}$ distribution above 150~GeV using an effective vertex function approach\cite{ZZaTGC}. Limits on aTGCs are set by obtaining 95\% confidence intervals for two CP-violating coupling parameters, $f_4^\gamma$ and $f_4^Z$, and two CP-conserving parameters, $f_5^\gamma$ and $f_5^Z$, and they are more stringent than corresponding limits from the $4\ell$ channel.

\subsection{$W^\pm W^\mp \rightarrow e^\pm \nu \mu^\mp \nu$}

Fiducial and differential cross-sections for $W^\pm W^\mp \rightarrow e^\pm \nu \mu^\mp \nu$ production \cite{WW} are measured using data recorded in 2015 and 2016. Events are required to contain exactly one electron and one muon of opposite charge and with $\pT>27$~GeV. They are required to have no additional leptons with $\pT>10$~GeV to reduce background from other diboson processes. The requirements of having no jets with $\pT>35$~GeV and no $b$-jets with $\pT>20$~GeV are imposed to suppress top-quark background.

The dominant background is from top-quark ($t\bar{t}$ and $Wt$) background, followed by non-prompt lepton background mainly due to $W$+jets, $Z\rightarrow\tau\tau$ and $WZ$. The purity of WW in selected events is 65\%. 

The cross-section is evaluated in the fiducial phase space, defined as $\pT^{\ell}>27$~GeV, $|\eta^\ell|<2.5$, $m(e\mu)>55$~GeV, $\pT^{e\mu}$, $E_\text{T}^\text{miss}>20$~GeV and no jets with $\pT>35$~GeV, $|\eta|<4.5$. The total uncertainty in the fiducial cross-section measurement is 7.1\%, dominated by $b$-tagging uncertainty, the jet energy scale uncertainty, and the modelling of the backgrounds. The integrated fiducial cross-section is measured to be $379\pm5~(\text{stat.})\pm27~(\text{syst.})$~fb, whereas the predicted cross-section at NNLO in pQCD obtained from \textsc{Matrix}, including NLO EW corrections and NLO in pQCD for the gluon-initiated production is $347\pm4~(\text{PDF})\pm19~(\text{scale})$~fb.

The fiducial cross-sections are also measured as a function of the jet-veto \pT-thresholds from 30~GeV in steps of 5~GeV up to 60~GeV. All predictions agree within uncertainties with the measurements but are consistently lower.

Six differential cross-sections are measured, as functions of $\pT^{\ell 1}$, the invariant mass of the dilepton system $m(e\mu)$, the transverse momentum of the dilepton system $\pT^{e\mu}$, the absolute rapidity of the dilepton system $|\text{y}(e\mu)|$, the difference in azimuthal angle between the two leptons $\Delta\phi(e, \mu)$ and $|\cos \theta^*|=|\tanh (\frac{\Delta\eta(e,\mu)}{2})|$. The unfolded $\pT^{\ell 1}$ distribution is used to study aTGCs, and limits are set on anomalous coupling parameters in an EFT framework\cite{EFT}.

\section{Electroweak production of diboson with two jets}

The EW production of diboson with two jets are searched for in multiple diboson final states, and has been observed in $W^\pm W^\pm$ \cite{ssWW}, $W^\pm Z$ \cite{VBSWZ} and $ZZ$ \cite{VBSZZ} with the ATLAS experiment. It is also searched for in $Z\gamma$ \cite{VBSZy} and in semileptonic decays of $WW/WZ/ZZ$ \cite{VBSsemi}, where no observation is yet made. Similar measurements are also performed by CMS where the production has been observed in $W^\pm W^\pm$ \cite{ssWWCMS} and $W^\pm Z$ \cite{VBSWZCMS}. 

To enhance the ratio of EW production to QCD-induced process and to select preferentially VBS process, certain characteristics of the VBS process are exploited. These events are characterized by a large invariant mass of the dijet system and a large rapidity separation of the two jets. The scattered quarks are not colour-connected and the hadronic activity between the two jets is expected to be low. The decay products of the bosons are also typically produced in the central region.

There is also interference between the SM electroweak and QCD-induced processes. The interference effect is estimated and is typically below 10\%. It is treated as a systematic uncertainty on the signal.

\subsection{$Z\gamma jj$}

The analysis \cite{VBSZy} uses data collected in 2015 and 2016. Events that contain a leptonically decaying $Z$ boson candidate, a photon and two jets are selected.

The main background comes from the QCD-induced production of the $Z\gamma jj$ final state, followed by $Z$+jets processes. $t\bar{t}\gamma$, $WZ$ and $Wt$ background are also considered but have smaller contributions.

Some of the event selection criteria defining the signal region are:
\begin{itemlist}
 \item Jet $\pT>50$~GeV, $N_{jets}\geq2$, pseudorapidity difference between the two leading jets $|\Delta\eta(j_1,j_2)|>1$, invariant mass of the dijet system $m(jj)>150$~GeV,
 \item $m(\ell\ell) + m(\ell\ell\gamma) > 182$~GeV, number of $b$-tagged jets $N_{b-jets}=0$ and $\zeta(\ell\ell\gamma)<5$, where $\zeta(\ell\ell\gamma)$ is the centrality of the $\ell\ell\gamma$ system relative to the tagging jets defined from the rapidity y of the $\ell\ell\gamma$ system and the two leading jets as

\begin{equation}
\zeta(\ell\ell\gamma) = \frac{\text{y}_{\ell\ell\gamma}-(\text{y}_{j_1}+\text{y}_{j_2})/2}{\text{y}_{j_1}-\text{y}_{j_2}}.
\end{equation}

\end{itemlist}

\begin{figure}[h]
\centerline{\includegraphics[width=2.5in]{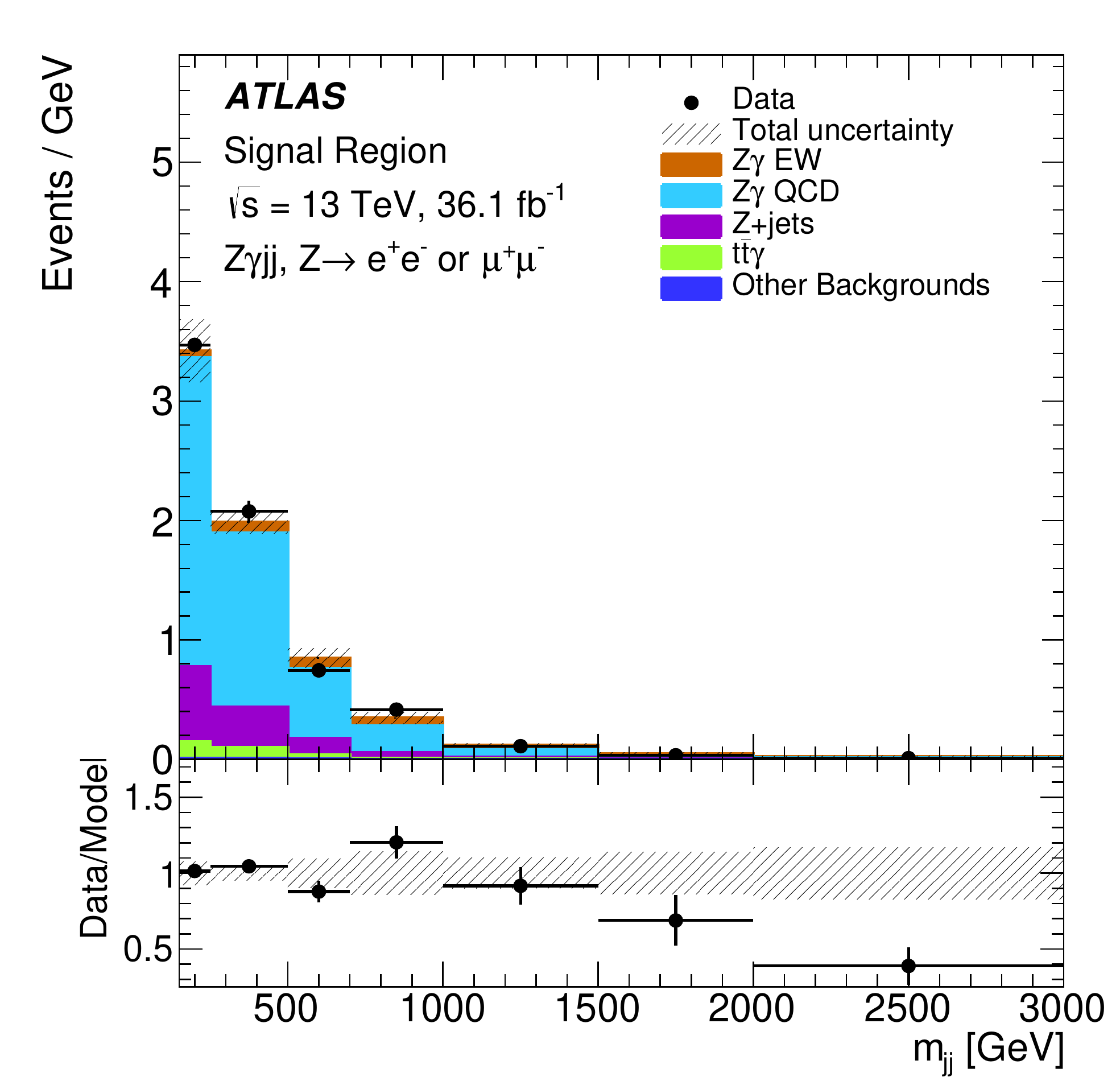}}
\vspace*{8pt}
\caption{Post-fit distribution of $m(jj)$ in the EW $Z\gamma jj$ signal region. The uncertainty band around the expectation includes all systematic uncertainties and takes into account their correlations as obtained from the fit. Reproduced from Ref.~\protect\refcite{VBSZy}.}
\protect\label{fig:VBSZy_mjjW}
\end{figure}

A boosted decision tree (BDT) trained using 13 kinematic variables is used to separate the EW signal from all the backgrounds. The modelling of the shapes of all input variables and their correlations by MC simulations is checked and good compatibility within the uncertainties is found except in the high mass tail of the $m(jj)$ distribution. This mismodelling is also observed in other analyses of EW processes such as $Zjj$ \cite{VBFZ8TeV, VBFZ} and $Wjj$ \cite{VBFW8TeV}. Figure \ref{fig:VBSZy_mjjW} illustrates the mismodelling in the high $m(jj)$.

The EW $Z\gamma jj$ signal significance and its fiducial cross-section are measured with a maximum likelihood to the BDT score distribution in the signal region. The observed EW $Z\gamma jj$ cross-section corresponds to the EW production plus the interference effect since the interference effect is not included in the QCD $Z\gamma jj$ contribution. The fiducial phase space is defined to closely follow the selection criteria of the signal region.

The signal strength is measured to be $1.00 \pm 0.19(\text{stat.}) \pm 0.13(\text{syst.})^{+0.13}_{-0.10}(\text{mod}.)$. Evidence for the EW $Z\gamma jj$ is reported with observed and expected significances of both 4.1~$\sigma$. The measured fiducial cross-section is $7.8 \pm 1.5(\text{stat.}) \pm 1.0(\text{syst.}) ^{+1.0}_{-0.8} (\text{mod.})~\text{fb}$, while the predicted LO fiducial cross-section from \textsc{MadGraph5}\_aMC@NLO 2.3.3 is $7.75 \pm 0.03(\text{stat.}) \pm 0.20(\text{PDF}+\alpha_\text{S}) + 0.40(\text{scale})~\text{fb}$.

\subsection{$ZZjj$}

The full run 2 dataset with an integrated luminosity of 139~$\text{fb}^{-1}$ is used to analyse the EW production of a $Z$ boson pair and two jets \cite{VBSZZ}. Both $ZZ\rightarrow\ell\ell\ell\ell$ and $ZZ\rightarrow\ell\ell\nu\nu$ final states originating from the decays of the $Z$ boson pair are considered. 

In the event selection, the two leading jets are required to be on opposite sides of the detector, i.e. $\text{y}_{j1} \times \text{y}_{j2}<0$. The requirement on the rapidity difference of the two jets $\Delta \text{y}(j_1,j_2)>2$ is imposed and $m(jj)$ is required to be greater than 300 (400) GeV in the $\ell\ell\ell\ell jj$ ($\ell\ell\nu\nu jj$) channel, where stricter selection in the $\ell\ell\nu\nu jj$ channel is optimized to suppress reducible backgrounds which are substantial in this channel.

In the $\ell\ell\ell\ell jj$ channel, the largest background arises from the QCD $ZZjj$ process, whose normalisation is constrained in a dedicated CR, defined by reversing either the $m(jj)$ or the $\Delta \text{y}(j_1,j_2)$ criteria. The impact on the extracted signal of a potential mismodelling of the $m(jj)$ in the QCD $ZZjj$ simulation is estimated by reweighting it in the SR using an $m(jj)$-dependent correction factor determined in a high centrality control region where the EW contribution is suppressed. The impact is found to be negligible. The normalisation of QCD $ZZjj$ production ($\mu_\text{QCD}^{\ell\ell\ell\ell jj}$) is varied simultaneously in the fit in the SR and QCD CR. In the $\ell\ell\nu\nu jj$ channel, the QCD $ZZjj$ process is modelled from simulation.

Similar to $Z\gamma jj$, multivariate discriminant based on BDT is used to separate the EW signal from background. From the combined channel, the observed $\mu_\text{EW}$ is $1.35\pm0.34$, while $\mu_\text{QCD}^{\ell\ell\ell\ell jj}$ is determined to be $0.96\pm0.22$. The dominant source of uncertainty is due to the limited number of data statistics. The background-only (no EW production) hypothesis is rejected with a statistical significance of $5.5~\sigma$ ($4.3~\sigma$ expected).

The EW $ZZjj$ cross-section in the combined fiducial volume in the $\ell\ell\ell\ell jj$ and $\ell\ell\nu\nu jj$ channels is found to be $0.82\pm0.21$~fb, calculated as $\mu_\text{EW}$ multiplied by the SM prediction of $0.61\pm0.03$~fb. In addition, the cross-sections for the production of inclusive $ZZjj$ are also measured. The measured cross-sections are $1.27\pm0.14$~fb for the $\ell\ell\ell\ell jj$ channel and $1.22\pm0.35$~fb for the $\ell\ell\nu\nu jj$ channel, both compatible with the SM predictions.

\subsection{$W^\pm W^\pm jj$}

The EW production of a same-sign $W$ boson pair with two jets is the most advantageous channel to observe such a process as it has the largest ratio of the EW to QCD production cross-sections. This study \cite{ssWW} uses 36 ~$\text{fb}^\text{-1}$ of collision data collected in 2015 and 2016.

The dominant background source is $WZ$, non-prompt lepton and electron charge mis-identification. QCD $W^\pm W^\pm jj$ production only constitutes a small background. 

The EW $W^\pm W^\pm jj$ cross-section is measured in a fiducial region defined as having exactly two same-sign leptons with $\pT>27$~GeV and $|\eta|<2.5$, $\Delta R (\ell,\ell)>0.3$, $m(\ell\ell)>20$~GeV, transverse momentum of the two neutrinos $\pT^{\nu\nu}>30$~GeV, two jets with leading (subleading) $\pT>65(35)$~GeV, $m(jj)>500$~GeV and $|\Delta \text{y} (j_1,j_2)|>2$.

Signal events are categorized into six mutually exclusive channels according to their lepton flavor and charge: $e^\pm e^\pm$, $e^\pm \mu^\pm$ and $\mu^\pm \mu^\pm$. The $m(jj)$  distributions in the signal (defined as $m(jj) > 500$~GeV) and control regions ($200<m(jj)<500$~GeV) are combined in a fit to extract the signal strength.

\begin{figure}[h]
\centerline{\includegraphics[width=2.5in]{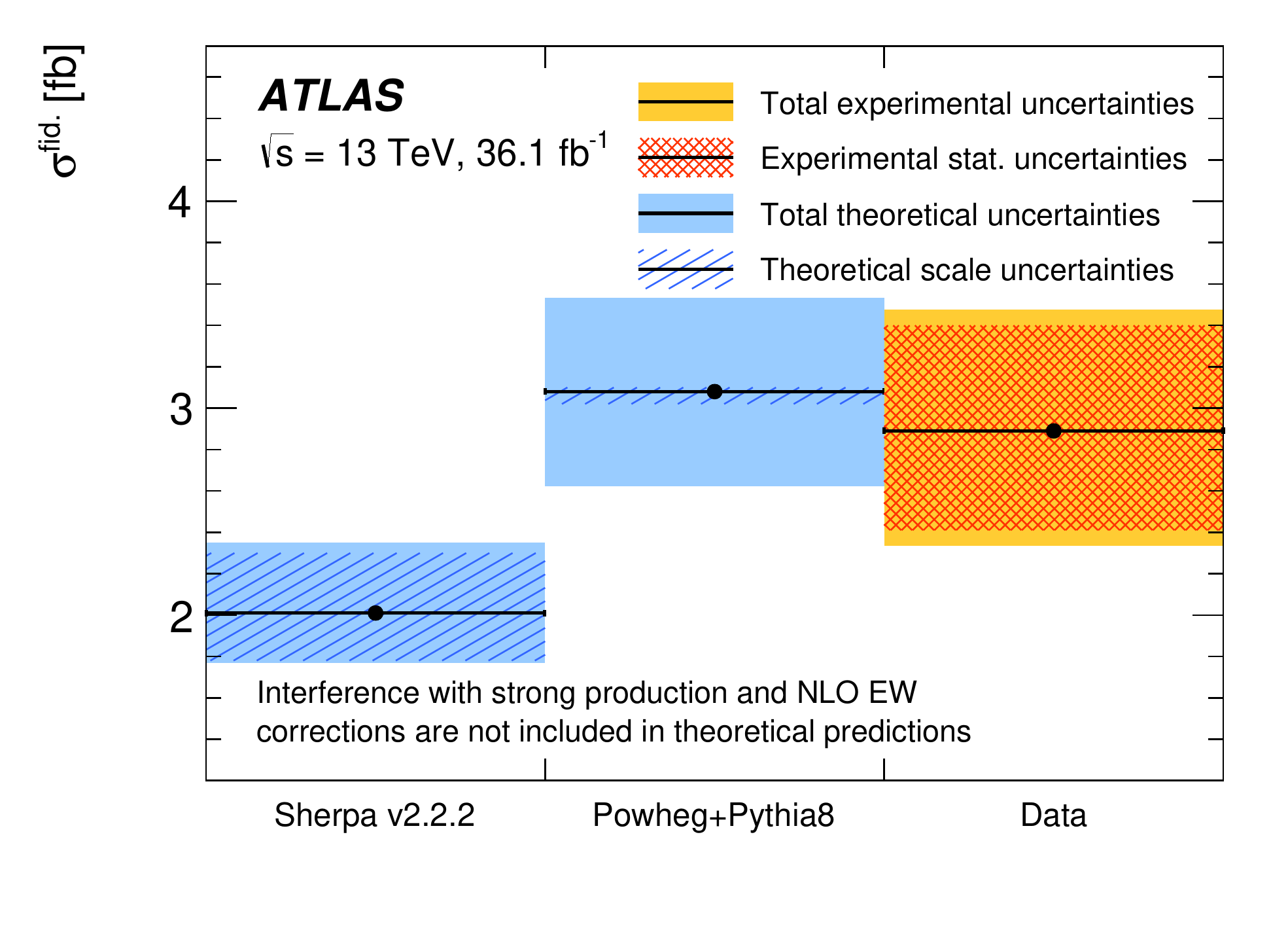}}
\vspace*{8pt}
\caption{Comparison of the measured EW $W^\pm W^\pm jj$ fiducial cross-section with theoretical predictions from \textsc{Sherpa} and \textsc{Powheg+Pythia8}. Reproduced from Ref.~\protect\refcite{ssWW}.}
\protect\label{fig:ssWW}
\end{figure}

The extracted signal strength is $1.44^{+0.26}_{-0.24}~(\text{stat.})^{+0.28}_{-0.22}~(\text{syst.})$, which is measured with respect to the \textsc{Sherpa} fiducial cross-section prediction. The observed signal significance is 6.5~$\sigma$. The dominant systematic uncertainty is due to backgrounds and jet energy and $E_\text{T}^\text{miss}$ scale and resolution. Figure \ref{fig:ssWW} shows the measured fiducial cross-section compared to the theoretical calculations from \textsc{Sherpa} v2.2.2 at LO in pQCD and \textsc{Powheg+Pythia8} at NLO, where the calculation from \textsc{Powheg+Pythia8} describes the measurement better.

\section{Summary}

The ATLAS collaboration has published a number of diboson measurements at $\sqrt{s}=13$~TeV, two of which use 139~$\text{fb}^{-1}$ of full Run 2 data. Both inclusive diboson measurements and searches for electroweak production in association with two jets which include vector boson scattering processes are studied. The results are generally in good agreement with the SM predictions at NNLO in pQCD in the case of inclusive diboson measurement and LO in pQCD in the case of electroweak production of $VVjj$. Sizeable higher-order corrections are seen in several processes. NLO EW corrections are also available for the inclusive diboson production and the effects can be quite large at high boson transverse momenta. The mismodelling of the dijet invariant mass of the QCD-induced $VVjj$ in simulation is observed.

\section*{Acknowledgments}

The author would like to thank Bogdan Malaescu for providing helpful comments on the draft.


\end{document}